\documentclass[12pt]{article}
\usepackage{amsmath}
\usepackage{color}
\usepackage{graphicx}
\begin{document}
\baselineskip=18 pt
\begin{center}
{\large{\bf Linear confinement of generalized KG-oscillator with a uniform magnetic field in Kaluza-Klein theory and Aharonov-Bohm effect }}
\end{center}

\vspace{.5cm}

\begin{center}
{\bf Faizuddin Ahmed}\\ 
{\bf National Academy, Gauripur-783331, Assam, India}\\
{\tt E-mail: faizuddinahmed15@gmail.com}
\end{center}

\vspace{.5cm}

\begin{abstract}

In this paper, we solve generalized KG-oscillator interacts with a uniform magnetic field in  five-dimensional space-time background produced by topological defects under a linear confining potential using the Kaluza-Klein theory. We solve this equation and analyze an analogue of the Aharonov-Bohm effect for bound states. We observe that the energy level for each radial mode depend on the global parameters characterizing the space-time, the confining potential, and the magnetic field which shows a quantum effect.

\end{abstract}

{\bf keywords:} Kaluza-Klein theory, Relativistic wave equation, electromagnetic interactions, Aharonov-Bohm effect, special functions.

\vspace{0.3cm}


\section{Introduction}

The ﬁrst proposal of a uniﬁed theory of fundamental interactions was elaborated by Kaluza \cite{Th} and Klein \cite{OZK} (see also, \cite{TM}). This new proposal established that the electromagnetism can be introduced through an extra (compactified) dimension in the space-time, where the spatial dimension becomes five-dimensional. This geometrical uniﬁcation of gravitation and electromagnetism in ﬁve-dimensional version of general relativity gave some interesting results. The idea behind introducing additional space-time dimensions has found wide applications in quantum field theory \cite{MBG}. Stationary cylindrically symmetric solutions to the ﬁve-dimensional Einstein and Einstein–Gauss–Bonnet equations has studied in \cite{MAA}. Few examples of these solutions are the ﬁve-dimensional generalizations of cosmic string, chiral cosmic string \cite{DG,PS2}, and magnetic ﬂux string \cite{MEX} space-times. 

The Kaluza-Klein theory (KKT) has investigated in several branches of physics. For example, in Khaler ﬁelds \cite{IMB}, in the presence of torsion \cite{GG,YSW}, in the Grassmannian context \cite{PE,RD,RD2}, in the description of geometric phases in graphene \cite{AY}, in Kaluza-Klein reduction of a quadratic curvature model \cite{SB}, in the presence of fermions \cite{DB,AM,SI}, and in studies of the Lorentz symmetry violation \cite{SMC,MG,APB}. In addition, the Kaluza-Klein theory has studied in the relativistic quantum mechanics, for example, the KG-oscillator on curved background in \cite{JC}, the KG-oscillator field interacts with Cornell-type potential in \cite{EHR}, generalized KG-oscillator in the background of magnetic cosmic string with scalar potential of Cornell-type in \cite{AHEP2}, generalized KG-oscillator in the background of magnetic cosmic string with a linear confining potential in \cite{EPJC2}, quantum dynamics of a scalar particle in the background of magnetic cosmic string and chiral cosmic string in \cite{CF,CF2}, bound states solution for a relativistic scalar particle subject to Coulomb-type potential in the Minkowski space-time in five dimensions in \cite{EVBL}, investigation of a scalar particle with position-dependent mass subject to a uniform magnetic field and quantum flux in the Minkowski space-time in five dimensions in \cite{EVBL2}, and quantum dynamics of KG-scalar particle subject to linear and Coulomb-type central potentials in the five-dimensional Minkowski space-time \cite{EVBL3}. Furthermore, eﬀects of rotation on KG-scalar ﬁeld subject Coulomb-type interaction and on KG-oscillator using Kaluza-Klein theory in the Minkowski space-time in five dimensions has also investigated \cite{EVBL4}. In order to describe singular behavior for a system at large distances in a uniformly rotating frame on the clocks and on a rotating body, Landau {\it et al.} \cite{LL} made a transformation such that it introduces a uniform rotation in the Minkowski space-time in cylindrical system. Non-inertial effects related to rotation have been investigated in several quantum systems, such as, in Dirac particle \cite{FWH}, on a neutral particle \cite{KB10}, on the Dirac oscillator \cite{PS20}, in cosmic string space-time \cite{LBC,LCNS}, in cosmic string space-time with torsion \cite{CJP}. Study of non-inertial effects on KG-oscillator within the Kaluza-Klein theory will be our next work.

The Klein-Gordon oscillator (KGO) \cite{SP,VVD} was inspired by the Dirac oscillator (DO) \cite{MM} applied to spin-$\frac{1}{2}$ particle. Several authors have studied KGO on background space-times, for example, in cosmic string, G\"{o}del-type space-times etc. ({\it e. g.} \cite{AB2,ZW,AHEP}). In the context of KKT, the KG-oscillator in five-dimensional cosmic string and magnetic cosmic string background \cite{JC}, under a Cornell-type potential in five-dimensional Minkowski space-time \cite{EHR} have investigated. In addition, generalized KGO on curved background space-time induced by a spinning cosmic string coupled to a magnetic field including quantum flux \cite{EPL}, in magnetic cosmic string background under the effects of Cornell-type potential \cite{AHEP2} and a linear confining potential \cite{EPJC2} using KKT, in the presence of Coulomb-type potential in $(1+2)$-dimensions G\"{u}rses space–time \cite{GERG}, in cosmic string space-time with a spacelike dislocation \cite{PS}, and in cosmic string space-time \cite{LFD} have investigated.

In this work, we study a generalized KGO by introducing a uniform magnetic field in the cosmic string line element using KKT \cite{Th,OZK,TM} under the effects of a linear confining potential, and analyze a relativistic analogue of the Aharonov-Bohm effect for bound states. The Aharonov-Bohm effect \cite{YA,MP,VBB,AB} is a quantum mechanical phenomena that describe phase shifts of the wave-function of a quantum particle due to the presence of a quantum flux produced by topological defects space-times. This effect has investigated by several authors in different branches of physics, such as, in Newtonian theory \cite{MAA2}, in bound states of massive fermions \cite{VRK}, in scattering of dislocated wave-fronts \cite{CC}, on position-dependent mass system under torsion effects \cite{CJP,AHEP,IJMPD}, in bound states solution of spin-$0$ scalar particles \cite{EPL}. In addition, this effect has investigated using KKT with or without interactions of various kind in five-dimensional the Minkowski or cosmic string space-time background \cite{JC,AHEP2,EPJC2,CF,CF2,EVBL,EVBL2,EVBL3,EVBL4,MPLA2}.

\section{Interactions of generalized KGO with scalar potential using the KKT }

The basic idea of the Kaluza-Klein theory \cite{Th,OZK,TM,CF} was to postulate one extra compactiﬁed space dimension and introducing pure gravity in new $(1+4)$-dimensional space-time. It turns out that the five-dimensional gravity manifests in our observable $(1 + 3)$-dimensional space-time as gravitational, electromagnetic and scalar ﬁled. In this way, we can work with general relativity in ﬁve-dimensions. The information about the electromagnetism is given by introducing a gauge potential $A_{\mu}$ in the space-time \cite{CF,CF2,EHR} as
\begin{equation}
ds^2=-dt^2+dr^2+\alpha^2\,r^2\,d\phi^2+dz^2+\left[ dy+\kappa\,A_{\mu} (x^{\mu})\, dx^{\mu} \right ]^2,
\label{1}
\end{equation}
where $\mu=0,1,2,3$, $x^0=t$ is the time-coordinate, $x^4=y$ is the coordinate associated with fifth additional dimension having ranges $0 < y < 2\,\pi\,a$ where, $a$ is the radius of the compact dimension of $y$, $(x^1=r, x^2=\phi, x^3=z)$ are the cylindrical coordinates with the usual ranges, and $\kappa$ is the gauge coupling or Kaluza constant \cite{CF}. The parameter $\alpha=(1-4\,\mu)$ \cite{AV} characterizing the wedge parameter where, $\mu$ is the linear mass density of the string. We assume the values of the parameter $\alpha$ lies in the range $0 < \alpha <1$.

Based on \cite{JC,EHR,AHEP2,CF,EVBL,EVBL2}, we introduce a uniform magnetic field $B_0$ and quantum flux $\Phi$ through the line-element of the cosmic string space-time (\ref{1}) in the following form
\begin{equation}
ds^2=-dt^2+dr^2+\alpha^2\,r^2\,d\phi^2+dz^2+\left[ dy+\left(-\frac{1}{2}\,\alpha\,B_0\,r^2+\frac{\Phi}{2\pi} \right)\, d\phi \right ]^2,
\label{2}
\end{equation}
where the gauge field given by
\begin{equation}
A_{\phi}=\kappa^{-1}\,\left (-\frac{1}{2}\,\alpha\,B_{0}\,r^2+\frac{\Phi}{2\pi} \right)
\label{3}
\end{equation}
gives rise to a uniform magnetic field $\vec{B}= \vec{\nabla}\times \vec{A}=-\kappa^{-1}\,B_0\,\hat{z}$ \cite{GAM}, $\hat{z}$ is the unitary vector in the $z$-direction. Here $\Phi=const$ is quantum flux \cite{YA,GAM} through the core of the topological defects \cite{CF3}.

The relativistic quantum dynamics of spin-$0$ scalar particle with a scalar potential $S(r)$ by modifying the mass term in the form $m \rightarrow m+ S(r)$ as done in \cite{AHEP2,EPJC2,AHEP} in five-dimensional case is described by \cite{EHR,EVBL,EVBL2,EVBL3}:
\begin{equation}
\left [\frac{1}{\sqrt{-g}}\,\partial_{M} (\sqrt{-g}\,g^{MN}\,\partial_{N})-(m+S)^2 \right]\,\Psi=0,
\label{4}
\end{equation}
where $M, N=0,1,2,3,4$, with $g=\det\,g=-\alpha^2\,r^2$ is the determinant of metric tensor $g_{MN}$ with $g^{MN}$ its inverse for the line element (\ref{2}) and $m$ is rest mass of the particle.

To couple generalized Klein-Gordon oscillator with field, following change in the radial momentum operator is considered \cite{EPL,AHEP2,EPJC2,GERG,PS,LFD,SZ}
\begin{equation}
\vec{p} \rightarrow \vec{p}-i\,m\,\Omega\,f(r)\hat{r}\quad \mbox{or}\quad \partial_{r} \rightarrow \partial_{r}+m\,\Omega\,f(r),
\label{oscillator}
\end{equation}
where $\Omega$ is the oscillator frequency and we can write $\vec{p}^{\,2} \rightarrow (\vec{p}+i\,m\,\Omega\,f(r)\hat{r})(\vec{p}-i\,m\,\Omega\,f(r)\hat{r})$. Therefore, the KG-equation becomes
\begin{equation}
\left [\frac{1}{\sqrt{-g}}\,\left (\partial_{M}+m\,\Omega\,X_{M} \right) \sqrt{-g}\,g^{MN}\,\left(\partial_{N}-m\,\Omega\,X_{N} \right)-(m+S)^2 \right]\,\Psi=0,
\label{oscillator2}
\end{equation}
where $X_{M}=(0, f(r), 0,0,0)$.

For the metric (\ref{2}) 
\begin{equation}
g^{MN}=\left (\begin{array}{lllll}
-1 & 0 & \quad 0 & 0 & \quad 0 \\
\quad 0 & 1 & \quad 0 & 0 & \quad 0 \\
\quad 0 & 0 & \quad \frac{1}{\alpha^2\,r^2} & 0 & -\frac{K\,A_{\phi}}{\alpha^2\,r^2} \\
\quad 0 & 0 & \quad 0 & 1 & \quad 0 \\
\quad 0 & 0 & -\frac{K\,A_{\phi}}{\alpha^2\,r^2} & 0 & 1+\frac{K^2\,A^2_{\phi}}{\alpha^2\,r^2}
\end{array} \right).
\label{5}
\end{equation}

By considering the line-element (\ref{2}) into the Eq. (\ref{oscillator2}), we obtain the following differential equation :
\begin{eqnarray}
&&[-\frac{\partial^2}{\partial t^2}+\frac{\partial^2}{\partial r^2}+\frac{1}{r}\,\frac{\partial}{\partial r}+\frac{1}{\alpha^2\,r^2}\,\left(\frac{\partial}{\partial \phi}-\kappa\,A_{\phi}\,\frac{\partial}{\partial y}\right)^2+\frac{\partial^2}{\partial z^2}+\frac{\partial^2}{\partial y^2}\nonumber\\
&&-m\,\Omega\,\left(f'+\frac{f}{r} \right)-m^2\,\Omega^2\,f^2(r)-\left (m+S \right)^2]\,\Psi (t,r,\phi,z,y)=0.
\label{6}
\end{eqnarray}
Since the line-element (\ref{2}) is independent of $t, \phi ,z, x$. One can choose the following ansatz for the function $\Psi$ as:
\begin{equation}
\Psi(t, r, \phi, z, y)=e^{i\,(-E\,t+l\,\phi+k\,z+q\,y)}\,\psi(r),
\label{7}
\end{equation}
where $E$ is the total energy of the particle, $l=0,\pm\,1,\pm\,2,.. \in {\bf Z}$, and $k, q$ are constants.

Substituting the ansatz (\ref{7}) into the Eq. (\ref{6}), we obtain the following equation:
\begin{eqnarray}
&&\left [\frac{d^2}{dr^2}+\frac{1}{r}\,\frac{d}{dr}+E^2-k^2-q^2-m\,\Omega\,\left(f'+\frac{f}{r} \right)- m^2\,\Omega^2\,f^2(r)-\frac{(l-K\,q\,A_{\phi})^2}{\alpha^2\,r^2}\right]\,\psi (r)\nonumber\\
&&=(m+S)^2\,\psi (r).
\label{8}
\end{eqnarray}

\vspace{0.2cm}
\subsection{ Linear confining potential}
\vspace{0.2cm}

In this work, we consider linear confining potential that studies in the confinement of quarks \cite{CLC}, in the relativistic quantum mechanics \cite{ZW,EPJC2,AHEP,HH,HH2,ME,MdM,SMI,SMI2}, and in atomic and molecular physics \cite{ICF}. This potential is given by
\begin{equation}
S (r)=\eta_L\,r
\label{9}
\end{equation}
where $\eta_L$ is a constant that characterizes the linear confining potential.

Below, we choose two types of function $f(r)$ for the studies of generalized KG-oscillator in the considered relativistic system subject to linear confining potential.

\vspace{0.2cm}
{\bf Case A \quad:\quad Cornell-type function $f (r)=b_1\,r+\frac{b_2}{r}$}
\vspace{0.2cm}

Substituting eqs. (\ref{3}) and (\ref{9}) into the Eq. (\ref{8}) and using the above function, we obtain the following equation:
\begin{equation}
\left [\frac{d^2}{dr^2}+\frac{1}{r}\,\frac{d}{dr}+\lambda-\frac{j^2}{r^2}-\omega^2\,r^2-b\,r \right]\,\psi (r)=0,
\label{10}
\end{equation}
where
\begin{eqnarray}
\lambda&=&E^2-k^2-q^2-m^2-2\,m\,\omega_c\,\frac{(l-\frac{q\,\Phi}{2\pi})}{\alpha}-2\,m\,\Omega\,b_1-2\,m^2\,\Omega^2\,b_1\,b_2,\nonumber\\
\omega&=&\sqrt{m^2\,\omega^2_{c}+\eta^2_{L}+m^2\,\Omega^2\,b^2_{1}},\nonumber\\
j&=&\sqrt{\frac{(l-\frac{q\,\Phi}{2\pi})^2}{\alpha^2}+m^2\,\Omega^2\,b^2_{2}},\nonumber\\
\omega_c&=&\frac{q\,B_0}{2\,m},\nonumber\\
b&=&2\,m\,\eta_L.
\label{11}
\end{eqnarray}

Introducing a new variable $\rho=\sqrt{\omega}\,r$, Eq. (\ref{10}) becomes
\begin{equation}
\left [\frac{d^2}{d\rho^2}+\frac{1}{\rho}\,\frac{d}{d\rho}+\zeta-\frac{j^2}{\rho^2}-\rho^2-\theta\,\rho \right]\,\psi (\rho)=0,
\label{12}
\end{equation}
where
\begin{equation}
\zeta=\frac{\lambda}{\omega}\quad,\quad \theta=\frac{b}{\omega^{\frac{3}{2}}}.
\label{13}
\end{equation}

Let us impose the requirement that the wave-function $\psi (\rho) \rightarrow 0$ for both $\rho \rightarrow 0$ and $\rho \rightarrow \infty$. Suppose the possible solution to Eq. (\ref{12}) is
\begin{equation}
\psi (\rho)=\rho^{j}\,e^{-\frac{1}{2}\,(\rho+\theta)\,\rho}\,H (\rho).
\label{14}
\end{equation}
Substituting the solution Eq. (\ref{14}) into the Eq. (\ref{12}), we obtain
\begin{equation}
H''(\rho)+\left [\frac{\gamma}{\rho}-\theta-2\,\rho \right ]\,H'(\rho)+\left [-\frac{\beta}{\rho}+\Theta \right]\,H (\rho)=0,
\label{15}
\end{equation}
where
\begin{eqnarray}
&&\gamma=1+2\,j,\nonumber\\
&&\Theta=\zeta+\frac{\theta^2}{4}-2\,(1+j),\nonumber\\
&&\beta=\frac{\theta}{2}\,(1+2\,j).
\label{16}
\end{eqnarray}
Equation (\ref{15}) is the biconfluent Heun's differential equation \cite{IJMPD,AHEP,EPJC2,AHEP2,AR,SYS} and $H (\rho)$ is the Heun polynomials.

The above equation (\ref{15}) can be solved by the Frobenius method. We consider the power series solution \cite{GBA}
\begin{equation}
H (\rho)=\sum_{i=0}^{\infty}\,c_{i}\,\rho^{i}
\label{17}
\end{equation}
Substituting the above power series solution into the Eq. (\ref{15}), we obtain the following recurrence relation for the coefficients:
\begin{equation}
c_{n+2}=\frac{1}{(n+2)(n+2+2\,j)}\,\left[\left\{\beta+\theta\,(n+1) \right\}\,c_{n+1}-(\Theta-2\,n)\,c_{n} \right].
\label{18}
\end{equation}
And the various coefficients are
\begin{eqnarray}
&&c_1=\frac{\theta}{2}\,c_0,\nonumber\\
&&c_2=\frac{1}{4\,(1+j)}\,[\left(\beta+\theta \right)\,c_{1}-\Theta\,c_{0}].
\label{19}
\end{eqnarray}

We must truncate the power series by imposing the following two conditions \cite{IJMPD,AHEP,EVBL,EVBL2,EVBL3, EPJC2,AHEP2}:
\begin{eqnarray}
\Theta&=&2\,n, \quad (n=1,2,...)\nonumber\\
c_{n+1}&=&0.
\label{21}
\end{eqnarray} 

By analyzing the condition $\Theta=2\,n$, we get the following second degree expression of the energy eigenvalues $E_{n,l}$:
\begin{eqnarray}
&&\frac{\lambda}{\omega}+\frac{\theta^2}{4}-2\,(1+j)=2\,n\nonumber\\\Rightarrow
&&E_{n,l}=\pm\,\{k^2+q^2+m^2+2\,\omega\,\left(n+1+\sqrt{\frac{(l-\frac{q\,\Phi}{2\pi})^2}{\alpha^2}+m^2\,\Omega^2\,b^{2}_2} \right)\nonumber\\
&&+2\,m\,\omega_c\,\frac{(l-\frac{q\,\Phi}{2\pi})}{\alpha}-\frac{m^2\,\eta^2_{L}}{\omega^2} +2\,m\,\Omega\,b_1\,(1+m\,\Omega\,b_2) \}^{\frac{1}{2}}.
\label{22}
\end{eqnarray}

Note that the Eq. (\ref{22}) does not represent the general expression for eigenvalues. One can obtain the individual energy eigenvalues one by one, that is, $E_1, E_2, E_3$,.. by imposing the additional recurrence condition $C_{n+1}=0$ on the eigenvalue as done in \cite{IJMPD,AHEP,EPJC2,AHEP2}. For $n=1$, we have $\Theta=2$ and $c_2=0$ which implies from Eq. (\ref{19})
\begin{eqnarray}
&&c_1=\frac{2}{\beta+\theta}\,c_0\Rightarrow \frac{\theta}{2}=\frac{2}{\beta+\theta}\nonumber\\
&&\omega_{1,l}=\left[\frac{b^2}{8}\,(3+2\,j) \right]^{\frac{1}{3}}= \left[\frac{m^2\,\eta^2_{1\,L}}{2}\,(3+2\,j) \right]^{\frac{1}{3}}
\label{23}
\end{eqnarray}
a constraint on the parameter $\omega_{1,l}$. The relation given in Eq. (\ref{23}) gives the  value of the parameter $\omega_{1,l}$ that permit us to construct a first degree polynomial solution of $H (\rho)$ for the radial mode $n=1$. Note that the parameter $\omega_{1,l}$ depends on the linear confining potential $\eta_L$ and its value changes for each quantum number $\{n, l\}$, so we have labeled $\omega \rightarrow \omega_{n,l}$ and $\eta_{L} \rightarrow \eta_{1\,L}$. Besides, we have adjusted the magnetic field $B^{1,l}_{0}$ and the linear confining potential $\eta_{1\,L}$ such that Eq. (\ref{23}) can be satisfied and we have simplified by labelling:
\begin{equation}
\omega^{1,l}_{c}=\frac{1}{m}\sqrt{\omega^2_{1,l}-\eta^2_{1\,L}-m^2\,\Omega^2\,b^2_{1}} \leftrightarrow B^{1,l}_{0}=\frac{2}{q}\sqrt{\omega^2_{1,l}-\eta^2_{1\,L}-m^2\,\Omega^2\,b^2_{1}}.
\label{24}
\end{equation}
It is noteworthy that the allowed value of the magnetic field $B^{1,l}_{0}$ for lowest state of the system given by (\ref{24}) is defined for the radial mode $n=1$. We can note from Eq. (\ref{24}) that the magnetic field $B_0$ depends on the quantum numbers $\{n, l\}$ of the relativistic system which shows a quantum effect. 

Therefore, the ground state energy level for $n=1$ is given by
\begin{eqnarray}
&&E_{1,l}=\pm\,\{k^2+q^2+m^2+2\,\omega_{1,l}\,\left(2+\sqrt{\frac{(l-\frac{q\,\Phi}{2\pi})^2}{\alpha^2}+m^2\,\Omega^2\,b^{2}_2} \right)\nonumber\\
&&+2\,m\,\omega^{1,l}_c\,\frac{(l-\frac{q\,\Phi}{2\pi})}{\alpha}-\frac{m^2\,\eta^2_{L}}{\omega^2_{1,l}}+2\,m\,\Omega\,b_1\,(1+m\,\Omega\,b_2) \}^{\frac{1}{2}}.
\label{25}
\end{eqnarray}
And the radial wave-functions is 
\begin{eqnarray}
\psi_{1,l}=\rho^{\sqrt{\frac{(l-\frac{q\,\Phi}{2\pi})^2}{\alpha^2}+m^2\,\Omega^2\,b^{2}_2}}\,e^{-\frac{1}{2}\,\left (\frac{2\,m\,\eta_L}{\omega^{\frac{3}{2}}_{1,l}}+\rho \right)\,\rho}\,\left(c_0+c_1\,\rho\right),
\label{26}
\end{eqnarray}
where
\begin{eqnarray}
c_1&=&\frac{1}{\sqrt{\frac{3}{2}+\sqrt{\frac{(l-\frac{q\,\Phi}{2\pi})^2}{\alpha^2}+m^2\,\Omega^2\,b^{2}_2}}}\,c_0.
\label{27}
\end{eqnarray}
Then, by substituting the magnetic field (\ref{24}) into the Eq. (\ref{25}), one can obtain the allowed values of the relativistic energy level for the radial mode $n=1$ of the system. As the values of the wedge parameter $\alpha$ are in the ranges $0 < \alpha < 1$, thus, the degeneracy of the energy is broken and shifted the energy level in comparison to the case of five-dimensional Minkowski space-time.

\vspace{0.5cm}
{\bf Case B\quad:\quad Coulomb-type function $f (r)=\frac{b_2}{r}$ }
\vspace{0.5cm}

In that case, the radial wave-equation (\ref{10}) becomes
\begin{equation}
\left [\frac{d^2}{dr^2}+\frac{1}{r}\,\frac{d}{dr}+\tilde{\lambda}-\frac{j^2}{r^2}-\tilde{\omega}^2\,r^2-b\,r \right]\,\psi (r)=0,
\label{b1}
\end{equation}
where
\begin{eqnarray}
\tilde{\lambda}&=&E^2-k^2-q^2-m^2-2\,m\,\omega_c\,\frac{(l-\frac{q\,\Phi}{2\pi})}{\alpha},\nonumber\\
\tilde{\omega}&=&\sqrt{m^2\,\omega^2_{c}+\eta^2_{L}}.
\label{b2}
\end{eqnarray}

Introducing a new variable $\rho=\sqrt{\tilde{\omega}}\,r$, Eq. (\ref{10}) becomes
\begin{equation}
\left [\frac{d^2}{d\rho^2}+\frac{1}{\rho}\,\frac{d}{d\rho}+\tilde{\zeta}-\frac{j^2}{\rho^2}-\rho^2-\tilde{\theta}\,\rho \right]\,\psi (\rho)=0,
\label{b3}
\end{equation}
where
\begin{equation}
\tilde{\zeta}=\frac{\tilde{\lambda}}{\tilde{\omega}}\quad,\quad \tilde{\theta}=\frac{b}{\tilde{\omega}^{\frac{3}{2}}}.
\label{b4}
\end{equation}

Let the possible solution to Eq. (\ref{b3}) is
\begin{equation}
\psi (\rho)=\rho^{j}\,e^{-\frac{1}{2}\,(\rho+\tilde{\theta})\,\rho}\,H (\rho).
\label{b5}
\end{equation}
Substituting solution Eq. (\ref{b5}) into the Eq. (\ref{b3}), we obtain
\begin{equation}
H''(\rho)+\left [\frac{\gamma}{\rho}-\tilde{\theta}-2\,\rho \right ]\,H'(\rho)+\left [-\frac{\tilde{\beta}}{\rho}+\tilde{\Theta} \right]\,H (\rho)=0,
\label{b6}
\end{equation}
where
\begin{eqnarray}
&&\tilde{\Theta}=\tilde{\zeta}+\frac{\tilde{\theta}^2}{4}-2\,(1+j),\nonumber\\
&&\tilde{\beta}=\frac{\tilde{\theta}}{2}\,(1+2\,j).
\label{b7}
\end{eqnarray}
Equation (\ref{b6}) is the biconfluent Heun's differential equation \cite{IJMPD,AHEP,EPJC2,AHEP2,AR,SYS} and $H (\rho)$ is the Heun polynomials.

Substituting the above power series solution (\ref{17}) into the Eq. (\ref{b6}), we obtain the following recurrence relation for the coefficients:
\begin{equation}
c_{n+2}=\frac{1}{(n+2)(n+2+2\,j)}\,\left[\left\{\tilde{\beta}+\tilde{\theta}\,(n+1) \right\}\,c_{n+1}-(\tilde{\Theta}-2\,n)\,c_{n} \right].
\label{b8}
\end{equation}
And the various coefficients are
\begin{eqnarray}
&&c_1=\frac{\tilde{\theta}}{2}\,c_0,\nonumber\\
&&c_2=\frac{1}{4\,(1+j)}\,[\left(\tilde{\beta}+\tilde{\theta} \right)\,c_{1}-\tilde{\Theta}\,c_{0}].
\label{b9}
\end{eqnarray}

We must truncate the power series by imposing the following two conditions \cite{IJMPD,AHEP,EVBL,EVBL2,EVBL3, EPJC2,AHEP2}:
\begin{eqnarray}
\tilde{\Theta}&=&2\,n, \quad (n=1,2,...)\nonumber\\
c_{n+1}&=&0.
\label{b10}
\end{eqnarray} 

By analyzing the condition $\tilde{\Theta}=2\,n$, we get the following second degree expression of the energy eigenvalues $E_{n,l}$:
\begin{eqnarray}
&&\frac{\tilde{\lambda}}{\tilde{\omega}}+\frac{\tilde{\theta}^2}{4}-2\,(1+j)=2\,n\nonumber\\\Rightarrow
&&E_{n,l}=\pm\,\{k^2+q^2+m^2+2\,\tilde{\omega}\,\left(n+1+\sqrt{\frac{(l-\frac{q\,\Phi}{2\pi})^2}{\alpha^2}+m^2\,\Omega^2\,b^{2}_2} \right)\nonumber\\
&&+2\,m\,\omega_c\,\frac{(l-\frac{q\,\Phi}{2\pi})}{\alpha}-\frac{m^2\,\eta^2_{L}}{\tilde{\omega}^2}\}^{\frac{1}{2}}.
\label{b11}
\end{eqnarray}

Following the similar technique as done earlier, we want to find the individual energy level and wave-function. For example $n=1$ we have $\Theta=2$ and $c_2=0$ which implies from Eq. (\ref{19})
\begin{eqnarray}
&&c_1=\frac{2}{\tilde{\beta}+\tilde{\theta}}\,c_0\Rightarrow \frac{\tilde{\theta}}{2}=\frac{2}{\tilde{\beta}+\tilde{\theta}}\nonumber\\
&&\tilde{\omega}_{1,l}=\left[\frac{b^2}{8}\,(3+2\,j) \right]^{\frac{1}{3}}
\label{b12}
\end{eqnarray}
a constraint on the parameter $\tilde{\omega}_{1,l}$. The magnetic field $B^{1,l}_{0}$ is so adjusted that Eq. (\ref{23}) can be satisfied and we have simplified by labelling:
\begin{equation}
\omega^{1,l}_{c}=\frac{1}{m}\sqrt{\tilde{\omega}^2_{1,l}-\eta^2_{L}} \leftrightarrow B^{1,l}_{0}=\frac{2}{q}\sqrt{\tilde{\omega}^2_{1,l}-\eta^2_{L}}.
\label{b13}
\end{equation}
We can see from Eq. (\ref{b13}) that the possible values of the magnetic field $B_0$ depend on the quantum numbers $\{n, l\}$ of the system as well as on the confining potential parameter.

Therefore, the ground state energy level for $n=1$ is given by
\begin{eqnarray}
&&E_{1,l}=\pm\,\{k^2+q^2+m^2+2\,\tilde{\omega}_{1,l}\,\left(2+\sqrt{\frac{(l-\frac{q\,\Phi}{2\pi})^2}{\alpha^2}+m^2\,\Omega^2\,b^{2}_2} \right)\nonumber\\
&&+2\,m\,\omega^{1,l}_c\,\frac{(l-\frac{q\,\Phi}{2\pi})}{\alpha}-\frac{m^2\,\eta^2_{L}}{\tilde{\omega}^2_{1,l}} \}^{\frac{1}{2}}.
\label{b14}
\end{eqnarray}
And the radial wave-functions is 
\begin{eqnarray}
\psi_{1,l}=\rho^{\sqrt{\frac{(l-\frac{q\,\Phi}{2\pi})^2}{\alpha^2}+m^2\,\Omega^2\,b^{2}_2}}\,e^{-\frac{1}{2}\,\left (\frac{2\,m\,\eta_L}{\tilde{\omega}^{\frac{3}{2}}_{1,l}}+\rho \right)\,\rho}\,\left(c_0+c_1\,\rho\right),
\label{b15}
\end{eqnarray}
where
\begin{eqnarray}
c_1&=&\frac{1}{\sqrt{\frac{3}{2}+\sqrt{\frac{(l-\frac{q\,\Phi}{2\pi})^2}{\alpha^2}+m^2\,\Omega^2\,b^{2}_2}}}\,c_0.
\label{b16}
\end{eqnarray}
Then, by substituting the real solution from Eq. (\ref{b12}) into the Eq. (\ref{b13}), it is possible to obtain the allowed values of the relativistic energy levels for the radial mode $n=1$ of the system. We can see that the lowest energy state defined by Eqs. (\ref{b12})--(\ref{b13}) plus the expression given in Eqs. (\ref{b14})--(\ref{b16}) is for the radial mode $n=1$, instead of $n=0$. This effect arises due to the presence of linear confining potential in the relativistic system. Since the wedge parameter $\alpha$ are in the ranges $0 < \alpha < 1$, thus, the degeneracy of the relativistic energy eigenvalue here also is broken and shifted the energy level in comparison to the case of five-dimensional Minkowski space-time.

\section{Conclusions}

In this work, we have investigated generalized Klein-Gordon oscillator with a uniform magnetic field subject to a linear confining potential in a topological defect five-dimensional space-time in the context of Kaluza-Klein theory. Linear confining potential has many applications such as confinement of quarks in particle physics and other branches of physics including relativistic quantum mechanics. For suitable total wave-function, we have derived the radial wave-equation for a Cornell-type function in {\it sub-section 2.1} and finally reached a biconfluent Heun's differential equation form. By power series method we have solved this equation and by imposing condition we have obtained the non-compact expression of the energy eigenvalues (\ref{22}). By imposing the recurrence condition $c_{n+1}=0$ for each radial mode, for example $n=11$, we have obtained the lowest state energy level and wave-function by Eqs. (\ref{25})--(\ref{27}), and others are in the same way. 

In {\it sub-section 2.2}, we have considered a Coulomb-type function on the same relativistic system with a linear confining potential. Here also we have reached a biconfluent Heun's equation form and following the similar technique as done earlier, we have obtained the non-compact expression of the energy eigenvalues (\ref{b11}). By imposing the additional recurrence condition $c_{n+1}=0$ on the eigenvalue, one can obtained the individual energy level and the corresponding wave-function, as for example, for the radial mode $n=1$ by Eqs. (\ref{b14})--(\ref{b16}), and others are in the same way. In {\it sub-section 2.1 –2.2}, we have seen that the presence of linear confining potential allow the formation of bound states solution of the considered relativistic system and hence, the lowest energy state is deﬁned by the radial mode $n=1$, instead of $n=0$. Also in gravitation and cosmology, the values of the wedge parameter $\alpha$ are in the ranges $0 < \alpha < 1$, and thus, the degeneracy of each energy level is broken and shifted the relativistic energy level in comparison to the case of five-dimensional Minkowski space-time. When we tries to analyze $c_{n+1}=0$ for the radial mode $n=1$, an observation is noted, where certain parameter is constraint, for example, $\omega_{1,l}$ that appears in Eq. (\ref{23}) in {\it sub-section 2.1 (Case A)} depend on the quantum number $\{n, l\}$ of the system as well as on linear confining potential $\eta_{L}$. Another interesting observation that we made in this work is the quantum eﬀect which arises due to the dependence of the magnetic ﬁeld $B^{n,l}_0$ on the quantum number $\{n, l\}$ of the relativistic system.

We have observed in this work that the angular momentum number $l$ of the system is shifted, $l \rightarrow l_0=\frac{1}{\alpha}\,(l-\frac{q\,\Phi}{2\,\pi})$, an effective angular quantum number. Therefore, the relativistic energy eigenvalues depends on the geometric quantum phase \cite{YA,GAM}. Thus, we have that, $E_{n, l} (\Phi+\Phi_0)=E_{n, l \mp \tau} (\Phi)$, where $\Phi_0=\pm\,\frac{2\pi}{q}\,\tau$ with $\tau=0,1,2,...$. This dependence of the relativistic energy level on the geometric quantum phase gives us a relativistic analogue of the Aharonov-Bohm effect for bound states.

The Kaluza-Klein theory lead to many new uniﬁed ﬁeld theories, for example, connecting this theory with supergravity resulted in improved supersymmetric Kaluza-Klein theory, multi-dimensional uniﬁed theories using the idea of Kaluza-Klein theory to postulate extra (compactiﬁed) space dimensions, superstring theory as well as the M-theory \cite{EW} model using the Kaluza-Klein theory in higher dimensions. The five-dimensional theory yields a geometrical interpretation of the electromagnetic field and electric charge. The relation of the five-dimensional KKT with the structure of fiber bundles was first made in \cite{AT} and the relationship between principal fiber bundles and higher dimensional theories is in \cite{YMC}.

\section*{Acknowledgement}

Author sincerely acknowledge the anonymous kind referee(s) for their value comments and suggestions which have greatly improved this present manuscript.

\section*{Conflict of Interest}

There is no potential conflict of interest regarding publication this manuscript. 

\section*{Contribution Statement}

Faizuddin Ahmed has done the whole work.

\end{document}